\documentclass[aps,prd,twocolumn,superscriptaddress,showpacs]{revtex4}

\usepackage{graphicx}

\usepackage{dcolumn}

\usepackage{bm}

\usepackage{epsfig}

\usepackage{amsmath}
\usepackage{color}
\usepackage{slashed}

\usepackage{mathrsfs}
\usepackage{amssymb}
\usepackage{float}

\usepackage{url}

\usepackage{tabularx}  

\begin{document}

\hfill FERMILAB PUB-13-083-E


\title{Measurement of R = \boldmath${\mathcal{B}\left( t \rightarrow Wb \right)/\mathcal{B}\left(t 
\rightarrow Wq \right)} $ in Top--quark--pair Decays using Lepton+jets Events and the Full CDF Run 
II Data set} 

\affiliation{Institute of Physics, Academia Sinica, Taipei, Taiwan 11529, Republic of China}
\affiliation{Argonne National Laboratory, Argonne, Illinois 60439, USA}
\affiliation{University of Athens, 157 71 Athens, Greece}
\affiliation{Institut de Fisica d'Altes Energies, ICREA, Universitat Autonoma de Barcelona, E-08193, Bellaterra (Barcelona), Spain}
\affiliation{Baylor University, Waco, Texas 76798, USA}
\affiliation{Istituto Nazionale di Fisica Nucleare Bologna, $^{ee}$University of Bologna, I-40127 Bologna, Italy}
\affiliation{University of California, Davis, Davis, California 95616, USA}
\affiliation{University of California, Los Angeles, Los Angeles, California 90024, USA}
\affiliation{Instituto de Fisica de Cantabria, CSIC-University of Cantabria, 39005 Santander, Spain}
\affiliation{Carnegie Mellon University, Pittsburgh, Pennsylvania 15213, USA}
\affiliation{Enrico Fermi Institute, University of Chicago, Chicago, Illinois 60637, USA}
\affiliation{Comenius University, 842 48 Bratislava, Slovakia; Institute of Experimental Physics, 040 01 Kosice, Slovakia}
\affiliation{Joint Institute for Nuclear Research, RU-141980 Dubna, Russia}
\affiliation{Duke University, Durham, North Carolina 27708, USA}
\affiliation{Fermi National Accelerator Laboratory, Batavia, Illinois 60510, USA}
\affiliation{University of Florida, Gainesville, Florida 32611, USA}
\affiliation{Laboratori Nazionali di Frascati, Istituto Nazionale di Fisica Nucleare, I-00044 Frascati, Italy}
\affiliation{University of Geneva, CH-1211 Geneva 4, Switzerland}
\affiliation{Glasgow University, Glasgow G12 8QQ, United Kingdom}
\affiliation{Harvard University, Cambridge, Massachusetts 02138, USA}
\affiliation{Division of High Energy Physics, Department of Physics, University of Helsinki and Helsinki Institute of Physics, FIN-00014, Helsinki, Finland}
\affiliation{University of Illinois, Urbana, Illinois 61801, USA}
\affiliation{The Johns Hopkins University, Baltimore, Maryland 21218, USA}
\affiliation{Institut f\"{u}r Experimentelle Kernphysik, Karlsruhe Institute of Technology, D-76131 Karlsruhe, Germany}
\affiliation{Center for High Energy Physics: Kyungpook National University, Daegu 702-701, Korea; Seoul National University, Seoul 151-742, Korea; Sungkyunkwan University, Suwon 440-746, Korea; Korea Institute of Science and Technology Information, Daejeon 305-806, Korea; Chonnam National University, Gwangju 500-757, Korea; Chonbuk National University, Jeonju 561-756, Korea; Ewha Womans University, Seoul, 120-750, Korea}
\affiliation{Ernest Orlando Lawrence Berkeley National Laboratory, Berkeley, California 94720, USA}
\affiliation{University of Liverpool, Liverpool L69 7ZE, United Kingdom}
\affiliation{University College London, London WC1E 6BT, United Kingdom}
\affiliation{Centro de Investigaciones Energeticas Medioambientales y Tecnologicas, E-28040 Madrid, Spain}
\affiliation{Massachusetts Institute of Technology, Cambridge, Massachusetts 02139, USA}
\affiliation{Institute of Particle Physics: McGill University, Montr\'{e}al, Qu\'{e}bec H3A~2T8, Canada; Simon Fraser University, Burnaby, British Columbia V5A~1S6, Canada; University of Toronto, Toronto, Ontario M5S~1A7, Canada; and TRIUMF, Vancouver, British Columbia V6T~2A3, Canada}
\affiliation{University of Michigan, Ann Arbor, Michigan 48109, USA}
\affiliation{Michigan State University, East Lansing, Michigan 48824, USA}
\affiliation{Institution for Theoretical and Experimental Physics, ITEP, Moscow 117259, Russia}
\affiliation{University of New Mexico, Albuquerque, New Mexico 87131, USA}
\affiliation{The Ohio State University, Columbus, Ohio 43210, USA}
\affiliation{Okayama University, Okayama 700-8530, Japan}
\affiliation{Osaka City University, Osaka 588, Japan}
\affiliation{University of Oxford, Oxford OX1 3RH, United Kingdom}
\affiliation{Istituto Nazionale di Fisica Nucleare, Sezione di Padova-Trento, $^{ff}$University of Padova, I-35131 Padova, Italy}
\affiliation{University of Pennsylvania, Philadelphia, Pennsylvania 19104, USA}
\affiliation{Istituto Nazionale di Fisica Nucleare Pisa, $^{gg}$University of Pisa, $^{hh}$University of Siena and $^{ii}$Scuola Normale Superiore, I-56127 Pisa, Italy, $^{mm}$INFN Pavia and University of Pavia, I-27100 Pavia, Italy}
\affiliation{University of Pittsburgh, Pittsburgh, Pennsylvania 15260, USA}
\affiliation{Purdue University, West Lafayette, Indiana 47907, USA}
\affiliation{University of Rochester, Rochester, New York 14627, USA}
\affiliation{The Rockefeller University, New York, New York 10065, USA}
\affiliation{Istituto Nazionale di Fisica Nucleare, Sezione di Roma 1, $^{jj}$Sapienza Universit\`{a} di Roma, I-00185 Roma, Italy}
\affiliation{Mitchell Institute for Fundamental Physics and Astronomy, Texas A\&M University, College Station, Texas 77843, USA}
\affiliation{Istituto Nazionale di Fisica Nucleare Trieste/Udine; $^{nn}$University of Trieste, I-34127 Trieste, Italy; $^{kk}$University of Udine, I-33100 Udine, Italy}
\affiliation{University of Tsukuba, Tsukuba, Ibaraki 305, Japan}
\affiliation{Tufts University, Medford, Massachusetts 02155, USA}
\affiliation{University of Virginia, Charlottesville, Virginia 22906, USA}
\affiliation{Waseda University, Tokyo 169, Japan}
\affiliation{Wayne State University, Detroit, Michigan 48201, USA}
\affiliation{University of Wisconsin, Madison, Wisconsin 53706, USA}
\affiliation{Yale University, New Haven, Connecticut 06520, USA}

\author{T.~Aaltonen}
\affiliation{Division of High Energy Physics, Department of Physics, University of Helsinki and Helsinki Institute of Physics, FIN-00014, Helsinki, Finland}
\author{S.~Amerio}
\affiliation{Istituto Nazionale di Fisica Nucleare, Sezione di Padova-Trento, $^{ff}$University of Padova, I-35131 Padova, Italy}
\author{D.~Amidei}
\affiliation{University of Michigan, Ann Arbor, Michigan 48109, USA}
\author{A.~Anastassov$^x$}
\affiliation{Fermi National Accelerator Laboratory, Batavia, Illinois 60510, USA}
\author{A.~Annovi}
\affiliation{Laboratori Nazionali di Frascati, Istituto Nazionale di Fisica Nucleare, I-00044 Frascati, Italy}
\author{J.~Antos}
\affiliation{Comenius University, 842 48 Bratislava, Slovakia; Institute of Experimental Physics, 040 01 Kosice, Slovakia}
\author{G.~Apollinari}
\affiliation{Fermi National Accelerator Laboratory, Batavia, Illinois 60510, USA}
\author{J.A.~Appel}
\affiliation{Fermi National Accelerator Laboratory, Batavia, Illinois 60510, USA}
\author{T.~Arisawa}
\affiliation{Waseda University, Tokyo 169, Japan}
\author{A.~Artikov}
\affiliation{Joint Institute for Nuclear Research, RU-141980 Dubna, Russia}
% ...

\collaboration{CDF Collaboration\footnote{With visitors from
$^a$University of British Columbia, Vancouver, BC V6T 1Z1, Canada,
$^b$Istituto Nazionale di Fisica Nucleare, Sezione di Cagliari, 09042 Monserrato (Cagliari), Italy,
$^c$University of California Irvine, Irvine, CA 92697, USA,
$^e$Institute of Physics, Academy of Sciences of the Czech Republic, 182~21, Czech Republic,
$^f$CERN, CH-1211 Geneva, Switzerland,
$^g$Cornell University, Ithaca, NY 14853, USA,
$^h$University of Cyprus, Nicosia CY-1678, Cyprus,
$^i$Office of Science, U.S. Department of Energy, Washington, DC 20585, USA,
$^j$University College Dublin, Dublin 4, Ireland,
$^k$ETH, 8092 Z\"{u}rich, Switzerland,
$^l$University of Fukui, Fukui City, Fukui Prefecture, Japan 910-0017,
$^m$Universidad Iberoamericana, Lomas de Santa Fe, M\'{e}xico, C.P. 01219, Distrito Federal,
$^n$University of Iowa, Iowa City, IA 52242, USA,
$^o$Kinki University, Higashi-Osaka City, Japan 577-8502,
$^p$Kansas State University, Manhattan, KS 66506, USA,
$^q$Brookhaven National Laboratory, Upton, NY 11973, USA,
$^r$University of Manchester, Manchester M13 9PL, United Kingdom,
$^s$Queen Mary, University of London, London, E1 4NS, United Kingdom,
$^t$University of Melbourne, Victoria 3010, Australia,
$^u$Muons, Inc., Batavia, IL 60510, USA,
$^v$Nagasaki Institute of Applied Science, Nagasaki 851-0193, Japan,
$^w$National Research Nuclear University, Moscow 115409, Russia,
$^x$Northwestern University, Evanston, IL 60208, USA,
$^y$University of Notre Dame, Notre Dame, IN 46556, USA,
$^z$Universidad de Oviedo, E-33007 Oviedo, Spain,
$^{aa}$CNRS-IN2P3, Paris, F-75205 France,
$^{cc}$Universidad Tecnica Federico Santa Maria, 110v Valparaiso, Chile,
$^{dd}$Yarmouk University, Irbid 211-63, Jordan,
$^{ll}$Universite catholique de Louvain, 1348 Louvain-La-Neuve, Belgium,
$^{oo}$University of Z\"{u}rich, 8006 Z\"{u}rich, Switzerland,
$^{pp}$Massachusetts General Hospital and Harvard Medical School, Boston, MA 02114 USA,
$^{qq}$Hampton University, Hampton, VA 23668, USA,
$^{rr}$Los Alamos National Laboratory, Los Alamos, NM 87544, USA
}}
\noaffiliation

\begin{abstract}
\noindent
We present a measurement of the ratio of the top-quark 
branching fractions 
$R=\mathcal{B}(t\rightarrow  Wb)/\mathcal{B}(t\rightarrow Wq)$, where $q$ represents quarks of type
 $b$, $s$, or  $d$, 
in the final state with a lepton and hadronic jets. 
The measurement uses $\sqrt{s}$ = 1.96 TeV proton--antiproton collision data from 8.7 fb$^{-1}$ 
of integrated luminosity
collected with the Collider Detector at Fermilab during Run II of the Tevatron. 
We simultaneously measure $R=0.94 \pm 0.09$ (stat+syst), the $t\bar{t}$ production cross section 
$\sigma_{t \bar t} = 7.5 \pm 1.0$ (stat+syst) pb. 
The magnitude of the Cabibbo-Kobayashi-Maskawa  matrix element, 
$\left|V_{tb}\right| = 0.97 \pm 0.05$ (stat+syst) is extracted 
assuming three generations of quarks,
and a lower limit of $|V_{tb}|>0.89$ at 95\% credibility level is set.

\end{abstract}

\maketitle                                                                                                                                      

In the standard model (SM) the top--quark decay rate into a $W$
boson and a down-type quark $q$ ($q = d, s, b$) is proportional to 
$\left|V_{tq}\right|^{2}$, the squared magnitude of the element of the Cabibbo-Kobayashi-Maskawa (CKM) 
matrix~\cite{CKM}. Under the assumption of a 3 $\times$ 3 unitary 
CKM matrix and using the existing constraints on $V_{ts}$ and $V_{td}$, the magnitude of the top--bottom quark coupling is
$\left |V_{tb}\right|=0.99915^{+0.00002}_{-0.00005}$ \cite{PDG, Vts}, and  the  
top quark decays almost exclusively to $Wb$ final states. 
Any significant deviation from the expected value would imply new physics: an extra generation of quarks, 
non-SM top-quark production, or non-SM background to top-quark production.
A direct measurement of the magnitude of the $V_{tb}$ matrix element can be 
obtained from    the single-top-quark production cross-section~\cite{singletop},
which is proportional to $\left|V_{tb}\right|^{2}$. The value of
$\left|V_{tb}\right|$ can also be extracted from the decay rate of pair-produced top quarks.
We define  $R$ as 
the ratio of the branching fractions
\begin{equation} 
R =\frac{\mathcal{B}(t\rightarrow Wb)}{\mathcal{B}(t\rightarrow Wq)}.
\label{eq:Rratio}
\end{equation}
Given the unitarity of the CKM matrix and assuming three generations,
$R$ is indirectly determined by the knowledge of 
$V_{ts}$ and $V_{td}$,
\begin{equation}
R = \frac{\left|V_{tb}\right|^2}{\left|V_{tb}\right|^2+\left|V_{ts}\right|^2+\left|V_{td}\right|^2}, 
\label{eq:Rratio2}
\end{equation}
and is derived to be $0.99830^{+0.00004}_{-0.00009}$~\cite{PDG}. A deviation from this prediction would be an 
indication of non-SM physics.

This Letter reports the first CDF simultaneous measurement of $R$ and 
top-quark-pair-production cross section
$\sigma_{t\bar{t}}$                                                       
performed on a data sample corresponding to an integrated luminosity of 8.7 fb$^{-1}$
 collected with the CDF II detector ~\cite{CDFII} at the Fermilab Tevatron $p\bar{p}$ Collider 
at center of mass energy $\sqrt{s}=1.96$ TeV. 
The analysis  uses events with a lepton and multiple jets in the final state,
where one $W$ boson coming from 
$t\bar t$ production  decays into a quark and an antiquark while the 
second $W$ boson decays into a charged lepton (electron or muon) and a neutrino. 

CDF has performed several
measurements of $R$ during both Run I and Run II, combining the lepton+jets
final state  with the dilepton final state, where both $W$ bosons decay into leptons. The most recent 
publication reported
$R=1.12^{+0.21}_{-0.19}$(stat) $^{+0.17}_{-0.13}$(syst), and $R>0.61$ at 95\% confidence level (C.L.) using
162 pb$^{-1}$ of integrated luminosity~\cite{R2005}.  The D0
Collaboration  measured  $R=0.90\pm0.04$ (stat+syst) and
$R>0.79$ at 95\% C.L.~\cite{d0meas} using data from 5.4 fb$^{-1}$ of integrated luminosity, 
in the lepton+jets and dilepton final states combined.  

The CDF II detector~\cite{CDFII} consists of a charged-particle
tracking system in a magnetic field of 1.4 T, segmented electromagnetic and hadronic calorimeters with a  pointing geometry, 
and muon detectors. A silicon microstrip detector provides determination of charged-particle trajectories 
(tracking) over the radial range 1.5 to 28 cm, and is essential
for the detection of displaced decay (secondary) vertices. 
A three-level, online event-selection system (trigger)~\cite{trigger} is used to select events with an 
electron (muon)  candidate in the central detector region
(pseudorapidity $| \eta | < 1.1 $)~\cite{coord}, 
with  $E_T$ ($p_{T}$) $>$ 18 GeV (18 GeV/$c$), 
which form the data set for this analysis. 

The measurement of $R$ is based on the determination of the number of $b$-quark jets in
$t \bar t$ events reconstructed in  the lepton+jets final state. 
The lepton+jets signature consists of a high-$p_{T}$ charged electron ($e$) or muon  ($\mu$),
large missing transverse energy $\slashed E_T$~\cite{coord} due to the undetected neutrino from the leptonic $W$ decay,
and at least three hadronic jets. Events containing  muons are classified according to the coverage of the
detectors used for their identification as central, when $| \eta | < 0.6$, and forward, when $0.6 < | \eta | < 1.0$.
Identification of jets coming from $b$-quark
fragmentation
 ($b$-jet tagging)
is performed by the {\sc{secvtx}}
algorithm, which is based on the reconstruction of secondary vertices displaced from the
primary $p\bar{p}$ interaction vertex and selects a sample enriched  with jets
originating from $b$ quarks~\cite{cuts}. 
The lepton+jets selection requirements are described  in Ref.~\cite{cuts}.
Briefly, the analysis requires the presence of one isolated lepton ($e$ or $\mu$)
with $E_T$ greater than 20 GeV,  $\slashed E_{T}$ of at least 20 GeV, and
 a minimum of three jets, reconstructed using
a cone algorithm~\cite{JES} with radius $\Delta R = 0.4 $ in $\eta - \phi$ space~\cite{coord},
within $\left|\eta\right|< 2.0$. The jet $E_{T}$, after correcting for the calorimeter response~\cite{JES},
has to exceed 30, 25, and 20 GeV for the most-energetic, second-most-energetic, and any additional jet
in the event, respectively.
The $W$-boson transverse mass~\cite{coord} is required to be greater than
20 GeV/$c^{2}$. Events with one  or two  identified $b$-jets are selected (1 $b$-tag and 2 $b$-tag events, respectively).

The background processes include $W$-boson production in association with heavy-flavor jets 
($Wb\bar{b}$, $Wc\bar{c}$, $Wc$), $W$-boson production in association 
with light-flavor jets that are incorrectly identified as $b$-jets (``mistags''), quantum chromodynamics 
 multijet (``QCD'') events containing misreconstructed or
real leptons or incorrectly-measured $\slashed E_T$, diboson events ($WW$, $WZ$, $ZZ$),  
single-top-quark production, and $Z$+jets events.

We divide the selected sample into subsets according to the type of lepton, number
of jets in the final state, and number of identified  $b$ jets (one or two).
As explained in more detail below, we derive an expected event yield for each category. We then maximize the
likelihood for observing the events found in each category 
by varying two fit parameters, $R$ and the top-quark-pair-production
cross section $\sigma_{t\bar{t}}$.

The $t\bar{t}$  events are modeled using the {\sc pythia}~\cite{pythia}  Monte Carlo (MC)
generator with top-quark mass $m_t$ = 172.5 GeV/$c^2$. 
We estimate the backgrounds
with a collection of data-driven and simulation techniques described in detail in Ref.~\cite{cuts}. 
The QCD background is modeled using data control samples~\cite{QCDback}. 
Mistags are estimated using a matrix (the mistag matrix) calculated in control samples and parametrized 
as a function of jet and event characteristics~\cite{QCDback}.
Diboson processes 
are simulated using {\sc pythia}; single-top-quark production is simulated by 
 {\sc powheg}~\cite{powheg}, while the parton shower and fragmentation is provided by {\sc pythia}.
The  {\sc alpgen}~\cite{alpgen} generator, with {\sc pythia} 
supplying the parton shower and fragmentation, is used to model the
$W$+jets and $Z$+jets backgrounds.
A {\sc geant}-based simulation is used to model the response of the CDF II
detector~\cite{geant}. The cross sections used for background normalization can be found 
in Ref.~\cite{QCDback}.
Table \ref{table:m2calc} shows the expected sample composition for all final states, after summing over lepton categories,
 assuming  $R=1$ and $\sigma_{t\bar{t}} =7.04\pm 0.49$ pb~\cite{topxsec}.

\begin{table*}[tb] 
\begin{center} 
\caption{Number of expected and observed events in lepton+jets data corresponding to 8.7 fb$^{-1}$ of integrated luminosity. 
Uncertainties include statistical and systematic contributions.}
\label{table:m2calc}
\begin{tabularx}{0.95\textwidth}{l@{\; \; \; \; \; \; \; \; }c@{\; }c@{\;
}c@{\; \; \; \; \; \; }c@{\; }c@{\; }c@{\; }c}

\hline \hline
              &\multicolumn{3}{c}{{ 1 $b$-tag}}&\multicolumn{3}{c}{{ 2 $b$-tags}}\\ 
\hline
Process &\multicolumn{1}{c}{{3 Jets}}&\multicolumn{1}{c}{{4 
Jets}}&\multicolumn{1}{c}{{$\geq $5 Jets}} &\multicolumn{1}{c}{{3 Jets}}&\multicolumn{1}{c}{{4 Jets}}&\multicolumn{1}{c}{{$\geq $5 Jets}}\\ 
\hline 
$t\bar t$ &800 $\pm$ 67 & 777 $\pm$ 64 & 260 $\pm$ 21 & 216 $\pm$ 30 & 271 $\pm$ 36 & 97 $\pm$ 13 \\ 
$W$+$b\bar{b}$ &291 $\pm$ 118 & 74 $\pm$ 30 & 17 $\pm$ 7& 48 $\pm$ 20 & 14 $\pm$ 6 & 4 $\pm$ 2\\ 
$W$+$c\bar{c}$ &167 $\pm$ 68 &47 $\pm$ 20 & 12 $\pm$ 5 & 5 $\pm$ 2 & 2 $\pm$ 1 & 0.8 $\pm$ 0.4 \\ 
$W$+$c$ &87 $\pm$ 35 &17 $\pm$ 7 & 4 $\pm$ 2 & 3 $\pm$ 1 & 0.8 $\pm$ 0.4 & 0.2 $\pm$ 0.1 \\ 
Single top &78 $\pm$ 7 & 17 $\pm$ 2 & 3.6 $\pm$ 0.3 & 18 $\pm$ 3 & 4.7 $\pm$ 0.7 & 1.1 $\pm$ 0.2\\
Diboson & 45 $\pm$ 5 & 11 $\pm$ 1 & 3.1 $\pm$ 0.3 & 3.1 $\pm$ 0.5 & 0.9 $\pm$ 0.1 & 0.30 $\pm$ 0.05 \\
$Z$+jets &32 $\pm$ 3 & 9.1 $\pm$ 0.9 & 2.4 $\pm$ 0.2 & 2.1 $\pm$ 0.2 & 0.77 $\pm$ 0.08 & 0.29 $\pm$ 0.03 \\
Mistags &303 $\pm$ 42 & 74 $\pm$ 14 & 17 $\pm$ 6 & 5 $\pm$ 1 & 1.7 $\pm$ 0.4 & 0.6 $\pm$ 0.2 \\ 
QCD &125 $\pm$ 50 & 35 $\pm$ 29 & 10 $\pm$ 9 & 6 $\pm$ 3 & 0.1 $\pm$ 1.5 & 0.1 $\pm$ 1.5 \\ \hline
Total prediction&1928 $\pm$ 243 & 1061 $\pm$ 93 & 330 $\pm$ 28 & 306 $\pm$ 40 & 296 $\pm$ 38 & 104 $\pm$ 13\\ 
Observed &1844 &1088 &339 & 275 & 273 & 126 \\ 
\hline \hline
\end{tabularx} 
\end{center} 
\end{table*}

The number of $b$-tagged events is the most sensitive quantity to possible values of $R$ different from one:
the smaller  $R$, the smaller  the probability to have a
$b$-jet in a top-quark-pair event.
Hence, the fraction of events with one or two tags is expected to decrease with
decreasing $R$.
In general, the $t\bar{t}$ production cross-section measured by CDF in the lepton+jets sample assumes $R$=1. 
In order to avoid any bias due to
this premise, we measure simultaneously $R$ and the production cross-section, since the measurement of the latter is 
affected by the sum of events in the different tag bins.

To perform the fit, we first divide the sample into 18 independent  subsamples, organized by type of lepton,
 number of jets in the event (3, 4, $\ge$ 5), and 
number of identified $b$-jets.
The expected number of  events in each  subsample, $\mu_{exp}^{i,j}$, 
is given by the following expression:
\begin{equation}
\mu_{exp}^{i,j} = \mu_{t\bar t}^{i,j} +N_B^{i,j} = \mathscr{L}^{j} \epsilon_{evt}^{i,j}  \sigma_{t\bar t} 
\epsilon_{tag}^i (R) + N_B^{i,j},
\label{eq:equationttexpected}
\end{equation}
where $\mu_{t\bar t}^{i,j}$ is the expected number of $t\bar t$ events and $N_B^{i,j}$ the expected 
number of background events. The  $i$ and $j$ indices indicate the $i$th jet bin  with one or two identified $b$-jets
(jet-tag category)  and the $j$th lepton category, respectively.  
$\mathscr{L}^{j}$ is the integrated luminosity, 
$\epsilon_{evt}^{i,j}$ includes the trigger and lepton identification efficiencies, and $\epsilon^{i}_{tag}(R)$ is 
the event-tagging efficiency, {\it i.e.}, the efficiency for tagging at least one jet in an event.
In an ideal case without background and assuming a $b$-tagging efficiency equal one for jets originating from $b$ quarks
and zero for
jets originating from non-$b$ quarks, the number of expected events with two  tags is proportional to $R^2$,
while the number of expected events
with one  tag is proportional to $2R(1-R)$.
The estimates for the background processes are calculated with various values of $R$. The differences with 
respect to the estimates obtained with $R=1$ are found to be negligible.


The event-tagging efficiencies 
are calculated in $t\bar{t}$ MC samples, using the probability to tag a jet as a $b$-jet according to the 
{\sc{secvtx}}
algorithm, on a jet-by-jet basis. For jets originated from $b$ and $c$ quarks, the $b$-jet 
tagging efficiencies are corrected for differences between data and MC using a scale factor 
SF = $0.96 \pm 0.05$~\cite{SF}. For jets originated from light-flavor quarks, the probability to tag them as 
$b$-jets is obtained using the mistag matrix.

In general, $\epsilon_{tag}^{i}$ is calculated from the  event probability 
to tag the $m$th event in MC processes with
possible $b$-quark final states. For an event with $n$ generic jets, the probability to have one 
(Eq. (\ref{eq:quattro})) or two
(Eq. (\ref{eq:cinque})) tagged
$b$-jets  becomes
\begin{equation}
P_{1-tag}^{m} = \sum_{q=1}^{n} p^{tag}_{q}   \left( \prod_{r=1,r\neq q}^{n} \left(1-p_{r}^{tag} \right) \right)
\label{eq:quattro}
\end{equation}
\begin{equation}
P_{2-tag}^{m} = \sum_{q=1}^{n-1} p^{tag}_{q}  \left(\sum_{r>q}^{n} p^{tag}_{r}  \left( \prod_{s=1,s\neq q, s\neq r}^{n} \left(1-p_{s}^{tag} \right) \right) \right),  
\label{eq:cinque}
\end{equation}
where $p_{q}^{tag}$ is either the probability to tag the $q$th jet, multiplied by the SF, for jets where the 
heavy-flavor quark is found inside the jet cone; or the mistag probability for jets matched to a light-flavor hadron, calculated using the mistag
matrix.

Finally, we use the $P_{l-tag}^m$ as an event  weight to calculate the event-tagging efficiency $\epsilon^i_{tag}$
for each subsample with $l$ tags and $n$ jets by summing the $P_{l-tag}^{m}$ weights over all of the pretagged events.

\begin{figure}
\begin{center}
\epsfig{file=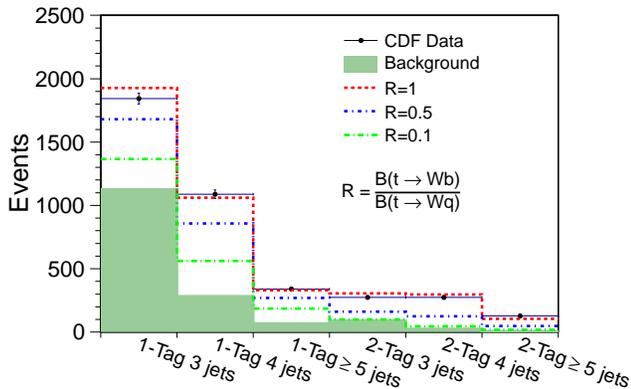,width=1.0\linewidth}
\caption{Observed events for the analysis final states after summing over lepton categories, 
compared to expected events for different values
of $R$. For the $t\bar{t}$ normalization the theoretical value  
 $\sigma_{t\bar{t}}=7.04\pm0.49$ pb is used.}
\label{fig:multipli}
\end{center}
\end{figure}

The MC sample employed for the $t\bar{t}$ signal modeling is generated assuming
 $\left|V_{tb}\right| = 1$ so it cannot be used 
directly to calculate  $\epsilon^{i}_{tag}$ as a function of $R$ through the 
algorithm described above. 
Instead, in the MC sample we assign 
a random number $P_{b}$ in the interval [0, 1] to every jet that is matched at the parton level to a $b$ quark 
from $t$-quark decay.
If $P_{b} <R$ we consider this jet as genuinely originated by a $b$ quark and use the tag probability multiplied by the SF as in
Eq. (\ref{eq:quattro}) and  (\ref{eq:cinque}); otherwise this jet is regarded as a light-flavor 
jet.
This simulates a configuration in which a $b$ quark produced in the top 
decay is a real $b$ only $R$ fraction of the time while  $(1-R)$ fraction of the time it is treated as a 
light-flavor quark and it is weighted by the mistag probability. This probabilistic approach allows the 
calculation of background and signal sample composition for any value of $R$.
This method
 reproduces exactly the standard calculation in the case of $R=1$, 
simulates $t\rightarrow Wq$ for $R=0$, and allows a calculation of 
$\epsilon^i_{tag}(R)$ 
in each tag subsample and in each jet bin.  
Figure~\ref{fig:multipli} shows the comparison of observed and expected events assuming $R$ equal to 1, 
0.5, and 0.1 in the various jet multiplicities and number of $b$-tagged jets in the final state.

In order to compare the prediction to the observed data in the 18 subsamples we use a 
likelihood function. We fit the observed event yields in each class of events to determine 
simultaneously $R$ and $\sigma_{t\bar t}$.
The likelihood function is:
\begin{equation}
\mathcal{L}=\prod_{i,j}\mathscr{P}\left(\mu^{i,j}_{exp}(R,\sigma_{t\bar t},x_{a})|N^{i,j}_{obs}\right)\prod_{a}G\left(x_{a}|0,1\right),
\end{equation}  
where $\mathscr{P}\left(\mu^{i,j}_{exp}(R,\sigma_{t\bar t},x_{a})|N^{i,j}_{obs}\right)$ is the Poisson probability to observe 
$N^{i,j}_{obs}$ events assuming the expected mean $\mu_{exp}^{i,j}$  (given by Eq. (\ref{eq:equationttexpected})), 
the index  $i$  indicates the jet-tag category, and the index  $j$ runs over the different lepton categories.
The estimates of the nuisance parameters $x_{a}$ are constrained to their central values and normalized to their 
uncertainties using Gaussian distributions  $G\left(x_{a}|0,1\right)$ centered at zero with unit variance.
This procedure takes into account correlations among channels by using same parameters for  
common sources of systematic uncertainties and allowing variations of each parameter  with respect to its central value. 

We perform the minimization of 
the negative logarithm of the likelihood $-2\log\left(\mathcal{L}\right)$, 
using the {\sc minuit} package~\cite{minuit}. We analitically extend $\epsilon^i_{tag}(R)$ beyond $R=1$ during the fitting procedure, 
constraining each individual $\epsilon^i_{tag}(R)$ to be greater than zero and their sum  to be $\leq 1$.
We simultaneously fit  $R$ and $\sigma_{t\bar t}$, which are the free parameters of the likelihood. 
We update the calculation of background yields using the value of $\sigma_{t\bar{t}}$ determined by the fit
and iterate the previous steps until the procedure converges.
No dependence on the starting point was observed in the results of the iterative procedure.

The uncertainty determined by the fit comprises the statistical contribution;  the systematic contribution on event-tagging efficiency, 
due to the systematic uncertainty on SF and the mistag matrix;  the event selection efficiency, 
due to the lepton-identification scale-factor and the trigger efficiency; 
 the background normalizations, 
including the heavy-flavor fractions; corrections for differences between MC and data heavy-flavor 
yields; and the luminosity~\cite{QCDback}.
We include separately the contributions due to the uncertainty on the jet-energy scale, 
effect of initial- and final-state radiation in the simulation (ISR/FSR), 
event-generator dependences, and top-quark mass.
The impact of the jet-energy scale uncertainty 
is estimated by varying the energy of all jets in the MC samples by  $\pm 1$ standard deviation 
with respect to the central value 
for both signal and backgrounds and by repeating the iterative fits.
The uncertainty arising from the choice of the MC generator is evaluated by repeating  the analysis
using a $t\bar t $ sample generated by {\sc herwig}~\cite{herwig}. 
The ISR/FSR uncertainty is evaluated by using $t\bar t $ MC samples generated with enhanced or suppressed 
radiation relative to the default configuration.
The theoretical value of the top-quark-production cross section depends on top-quark mass~\cite{masscrosscalc1}.
The recursive fit of $\sigma_{t\bar t}$ is expected to reduce 
the impact of this systematic uncertainty.
In order to check this assumption, we repeat the measurement 
using two different MC samples 
for the $t \bar t$ signal, simulated with $m_{t}=170$ GeV/$c^{2}$ and $m_{t}=175$ GeV/$c^{2}$, respectively. 
Central values and uncertainties on those systematic effects
 are included in the likelihood as nuisance parameters.

As a consistency check, the effect of each  source of systematic uncertainty is estimated via simulated 
experiments. For each source  we generate a set of simulated experiments with the same 
prescription but with the nuisance parameter $x_{a}$, relative to the 
systematic effect under study, shifted by one standard deviation from its nominal value.
 We determine the effect of changing each source of systematic uncertainty as 
 the change in the mean of the distributions of  $R$ 
and $\sigma_{t\bar{t}}$. 
Table \ref{table:systematics} lists the various systematic uncertainties assumed as fully uncorrelated.

\begin{table}
\caption{Systematic uncertainties on the measurement of $R$ and $\sigma_{t\bar{t}}$
obtained from simulated experiments. ``Others'' indicates the squared sum of minor systematic uncertainties.
All systematic uncertanties are assumed to be fully uncorrelated. The statistical uncertainty is shown as well.}
\begin{tabular}{lcccc}
\hline \hline
Source     &$+\delta R$&$-\delta R$&$+\delta\sigma_{t\bar{t}}$&$-\delta\sigma_{t\bar{t}}$\\
                         &           &           & (pb)      &    (pb)\\
\hline
$b$-tagging      &0.078  &--0.073  &0.06  &--0.03\\
Background normalization&0.056&--0.052&0.78&--0.66\\
Jet-energy scale & 0.016 & --0.019 & 0.46 &--0.41\\
ISR/FSR          &0.006  & --0.006 &0.22  &--0.21\\
Luminosity&0.001&--0.002&0.44&--0.39\\
Top-quark mass&0.001&--0.000&0.33&--0.32\\
Others & 0.005 & --0.006 & 0.17 & --0.15 \\
\hline
Total syst. uncert. &   0.088 & --0.081 & 1.04 & --0.92 \\
\hline
Statistical&                0.043    &   --0.043    &     0.29     &  --0.29         \\
\hline
\hline
\end{tabular}
\centering
\label{table:systematics}
\end{table}

\begin{figure}[tbh]
\begin{center}
\epsfig{file=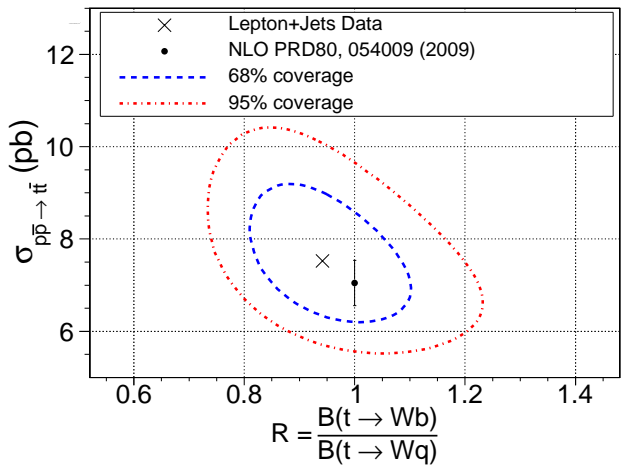,width=1.0\linewidth}
\caption{The fit results for the simultaneous measurement of $R$ and $\sigma_{ t\bar{t}}$.
The X-cross corresponds to the  maximum of the likelihood; the point with error bar to the NLO cross section calculation. 
The two-dimensional confidence regions are  shown as well.}
\label{fig:contorni}
\end{center}
\end{figure}

The final results 
are  summarized in Table \ref{table:resultssum}. 
 Figure \ref{fig:contorni}  shows the two-dimensional likelihood contour in the 
($R$, $\sigma_{t\bar{t}}$) plane, for the  fit including statistical and systematic 
uncertainties. 
The best-fit value is indicated with a ``X'' and can be 
compared to the theoretical SM prediction at the next-to-leading (NLO) order expansion in the strong-interaction 
coupling constant~\cite{topxsec}. 
The results are in agreement with the theoretical prediction to within one standard deviation.

\begin{figure}[h!]
\begin{center}
\epsfig{file=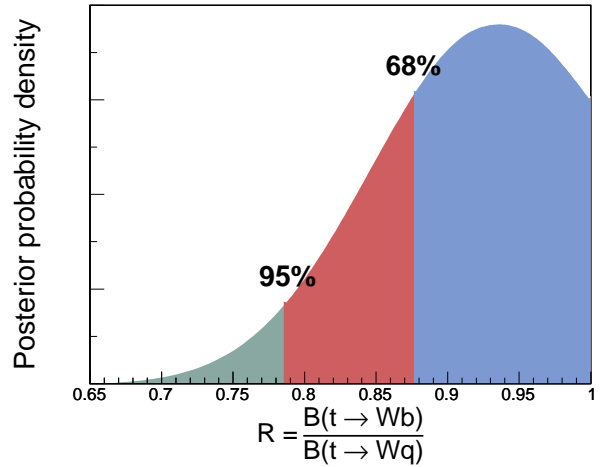,width=1.0\linewidth}
\caption{Lower limits on $R$ at various credibility levels obtained by the integration of the posterior probability.}
\label{fig:fitmargin}
\end{center}
\end{figure}

To determine the credibility level limit on $R$ 
we follow a Bayesian statistical approach. Since $R$ is bounded to be in 
the interval [0,1], the prior probability density 
is chosen to be zero outside these $R$ boundaries, 
while we consider all physical values equally probable. To obtain the posterior distribution for $R$, 
we integrate over all nuisance parameters using non-negative normal distributions as priors. 
We also integrate over $\sigma_{t\bar{t}}$
with the only constraint to be positive defined. 
The Bayesian lower limits at 
$68$\% and $95$\% credibility levels  are shown in 
Fig.~\ref{fig:fitmargin} and yield $R>0.785$ at 95\% c.l. 
From Eq. (\ref{eq:Rratio}) we extract a measurement of $V_{tb}$. Assuming 
three generations of quarks and given the unitarity of the CKM matrix, we have 
$\left|V_{td}\right|^{2}+\left|V_{ts}\right|^{2}+\left|V_{tb}\right|^{2}=1$,
leading to $R=\left|V_{tb}\right|^{2}$. 
From the fit results we obtain $\left|V_{tb}\right|=0.97\pm0.05$ and 
$\left|V_{tb}\right|>0.89$ at 95\% credibility levels.

 \begin{table}[thb]
\caption{Measured values resulting from the simultaneous likelihood fit of
$\sigma_{t\bar{t}}$ and $R$. The uncertainties on $R$ and $\sigma_{t\bar{t}}$ correspond to a variation of one  unit 
of $-2\log\left(\mathcal{L}\right)$. The correlation parameter is 
$\rho = - 0.434$. The magnitude of the CKM matrix element $\left|V_{tb}\right|$  is 
derived from $R=\left|V_{tb}\right|^{2}$. Lower limits at different 
credibility levels (c.l.) are obtained by
 integration of the posterior probability distribution.}
 \centering
 \begin{tabular}{cccc}
 \hline \hline
 Parameter                                   & Value          & Lower limit           & Lower limit  \\
                                             & (stat+syst)                          
&     68\% c.l.         &  95\% c.l.  \\
 \hline
 $\sigma_{t\bar{t}}$ (pb) & $7.5 \pm 1.0 $  &       -             &       -            \\
 $R$                                            & $0.94 \pm 0.09$ &      0.876          &      0.785         \\
 $\left|V_{tb}\right|$                        & $0.97 \pm 0.05$ &      0.936          &      0.886         \\
 \hline \hline
 \end{tabular}
 \label{table:resultssum}
 \end{table}

In summary, we present the simultaneous measurement of 
$R = 0.94 \pm 0.09$ and  $\sigma_{t\bar{t}} = 7.5 \pm 1.0$ pb with a 
correlation $\rho = -0.434$, 
and the determination of  $\left|V_{tb}\right| = 0.97 \pm 0.05$.  
The results for $R$ and $\left|V_{tb}\right|$ are the most precise determination obtained by CDF and 
are in agreement with the standard model~\cite{PDG}, 
with the previous CDF measurements~\cite{R2005}, with the latest
measurement of $R$ performed by D0~\cite{d0meas}, 
and with the direct  
measurement of single-top-quark production cross section 
 performed by LHC~\cite{singleLHC}  and Tevatron~\cite{singleTeV,singletop} experiments.

We thank the Fermilab staff and the technical staffs of the participating institutions for their vital 
contributions. This work was supported by the U.S. Department of Energy and National Science Foundation; 
the Italian Istituto Nazionale di Fisica Nucleare; the Ministry of Education, Culture, Sports, Science 
and Technology of Japan; the Natural Sciences and Engineering Research Council of Canada; the National 
Science Council of the Republic of China; the Swiss National Science Foundation; the A.P. Sloan Foundation; 
the Bundesministerium f\"ur Bildung und Forschung, Germany; the Korean World Class University Program, 
the National Research Foundation of Korea; the Science and Technology Facilities Council and the Royal Society, 
UK; the Russian Foundation for Basic Research; the Ministerio de Ciencia e Innovaci\'{o}n, and Programa 
Consolider-Ingenio 2010, Spain; the Slovak R\&D Agency; the Academy of Finland; and the Australian 
Research Council (ARC); and the EU community Marie Curie Fellowship contract 302103.

\nocite{*}
\appendix


\begin{thebibliography}{150}
\bibitem{coord} We use a cylindrical coordinate system 
where the $z$ axis is along the proton beam direction, $\phi$ is the azimuthal angle, and 
$\theta$ is the polar angle. Pseudorapidity is $\eta = -\ln\tan(\theta/2)$, 
while transverse momentum is $p_{T}=\left|p\right|\sin\theta$, and transverse energy is $E_{T}=E\sin\theta$. 
Missing transverse energy, $\slashed{E}_{T}$, is defined as the magnitude of $-\sum_{i}E_{T}^{i}\hat{n}_{i}$, 
where $\hat{n}_{i}$ is the unit vector in the azimuthal plane 
that points from the beam line to the $i$th calorimeter tower.
The $W$ boson transverse mass is defined as
$M_{T}^{W} = \frac{1}{c^{2}}\sqrt{2 E_{T}^{l}\slashed E_{T}\left(1-\cos\phi_{l\nu}\right)}$,
where $E_{T}^{l}$ is the transverse energy of the lepton and $\phi_{l\nu}$ is the angle between the lepton and the $\slashed E_{T}$.
\end{thebibliography}
\end{document}